\documentclass[10pt,english,sigconf]{acmart}
\pdfoutput=1
\usepackage[scaled=0.9]{beramono}
\usepackage[T1]{fontenc}
\usepackage[latin9]{inputenc}
\usepackage{prettyref}
\usepackage{booktabs}
\usepackage{enumitem}
\usepackage{graphicx}

\makeatletter

\providecommand{\tabularnewline}{\\}

\usepackage{color}
\usepackage{listings}
\usepackage{epsfig,endnotes}
\usepackage[shortcuts]{extdash} %
\usepackage{siunitx} %
\usepackage{csvsimple}
\usepackage{balance}

\usepackage{prettyref}
\newrefformat{cha}{\hyperref[#1]{Chapter~\ref*{#1}}}
\newrefformat{sec}{\hyperref[#1]{Sec.~\ref*{#1}}}
\newrefformat{sub}{\hyperref[#1]{Sec.~\ref*{#1}}}
\newrefformat{subsec}{\hyperref[#1]{Sec.~\ref*{#1}}}
\newrefformat{tab}{\hyperref[#1]{Tab.~\ref*{#1}}}
\newrefformat{fig}{\hyperref[#1]{Fig.~\ref*{#1}}}
\newrefformat{lst}{\hyperref[#1]{Listing~\ref*{#1}}}
\newrefformat{pat}{\hyperref[#1]{Patch~\ref*{#1}}}
\newrefformat{alg}{\hyperref[#1]{Algorithm~\ref*{#1}}}
\newrefformat{eq}{\hyperref[#1]{Eq.~\ref*{#1}}}

\usepackage{xcolor}
\definecolor{darkgrey}{rgb}{.4, .4, .4}
\definecolor{lightgrey}{rgb}{.95, .95, .95}
\definecolor{darkgreen}{rgb}{0, .5, 0}
\definecolor{darkred}{rgb}{.5, 0, 0}
\definecolor{lightred}{rgb}{1, .5, .5}

\usepackage{listings}
\lstdefinelanguage{hsclinebreaks}{
literate={\_}{}{0\discretionary{\_}{}{\_}} {/}{}{0\discretionary{/}{}{/}}%
}

\usepackage{pgfplots}
\usepackage{pgfplotstable}
\pgfplotsset{compat = 1.14}

\usetikzlibrary{plotmarks}
\usetikzlibrary{decorations.pathreplacing,calc}

\author{Alexander Lochmann}
\affiliation{%
  \institution{TU Dortmund}
  \department{Department of\\Computer Science 6}
  \streetaddress{Otto-Hahn-Str. 14}
  \city{Dortmund}
  \country{Germany}
  \postcode{44227}
}
\author{Robin Thunig}
\affiliation{%
  \institution{TU Dortmund}
  \department{Department of\\Computer Science 12}
  \streetaddress{Otto-Hahn-Str. 16}
  \city{Dortmund}
  \country{Germany}
  \postcode{44227}
}
\author{Horst Schirmeier}
\affiliation{%
  \institution{TU Dortmund}
  \department{Department of\\Computer Science 12}
  \streetaddress{Otto-Hahn-Str. 16}
  \city{Dortmund}
  \country{Germany}
  \postcode{44227}
}

\copyrightyear{2020}
\renewcommand\@copyrightpermission{\begin{minipage}{0.5\columnwidth}\href{https://creativecommons.org/licenses/by-sa/4.0/}{\includegraphics{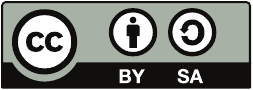}}\end{minipage}
\hspace{\fill}\begin{minipage}{0.45\columnwidth}Except as otherwise noted, this paper is licenced under the Creative Commons Attribution-Share Alike 4.0 International Licence.\end{minipage}}
\renewcommand\@copyrightowner{Copyright held by the authors.}
\acmConference[FGBS '20]{Herbsttreffen Fachgruppe Betriebssysteme}{September 24--25, 2020}{Aaachen, Germany}
\acmPrice{}
\acmISBN{}
\acmDOI{10.18420/fgbs2020h-01}
\makeatother

\usepackage{babel}
\usepackage{listings}

\begin{document}
\title[Improving Linux-Kernel Tests for LockDoc with Feedback-driven Fuzzing]{Improving Linux-Kernel Tests for LockDoc\\
with Feedback-driven Fuzzing}

\begin{abstract}
LockDoc is an approach to extract locking rules for kernel data structures
from a dynamic execution trace recorded while the system is under
a benchmark load. These locking rules can e.g. be used to locate synchronization
bugs. For high rule precision and thorough bug finding, the approach
heavily depends on the choice of benchmarks: They must trigger the
execution of as much code as possible in the kernel subsystem relevant
for the targeted data structures. However, existing test suites such
as those provided by the \emph{Linux Test Project} (LTP) only achieve
-- in the case of LTP -- about 35 percent basic-block coverage for
the VFS subsystem, which is the relevant subsystem when extracting
locking rules for filesystem-related data structures.

In this article, we discuss how to complement the LTP suites to improve
the code coverage for our LockDoc scenario. We repurpose \emph{syzkaller}
-- a coverage-guided fuzzer with the goal to validate the robustness
of kernel APIs -- to 1)~\emph{not} aim for kernel crashes, and to
2)~maximize code coverage for a specific kernel subsystem. Thereby,
we generate new benchmark programs that can be run in addition to
the LTP, and increase VFS basic-block coverage by 26.1 percent.
\end{abstract}
\maketitle

\section{Introduction\label{sec:Introduction}}

Over a period of more than a decade, the Linux kernel underwent a
transformation from Linux 2.0's coarse-grained and sturdy Big Kernel
Lock \citep{Bovet:2005:ULK:1077084} to more and more fine-grained
synchronization on the granularity of kernel subsystems and even single
data structures \citep{Bovet:2005:ULK:1077084,Love:2010:LKD:1855096,lockingguide}.
While reducing lock contention and scaling better to multi- and many-core
platforms, fine-grained locking is error-prone and has led to numerous
synchronization bugs in the past. This situation is exacerbated by
incomplete, inconsistent and in parts faulty locking documentation.

Our LockDoc approach \citep{lochmann:19:lockdoc} addresses these
issues by tracing locking patterns and data-structure accesses in
a running Linux kernel under load, and deriving locking rules --
i.e., which locks in which particular order must be taken to access
a specific data-structure element -- from this trace. The derived
locking rules can consequently be used to validate or generate documentation,
and to locate synchronization bugs. However, locking-rule quality
and the capability to find bugs also in remote parts of the kernel
heavily depends on how the system is put under load -- i.e. on the
choice of benchmark programs that trigger the execution of kernel
code by invoking system calls. Focusing on the Linux kernel's \emph{Virtual
File System} (VFS) subsystem in our LockDoc study \citep{lochmann:19:lockdoc},
we relied on a filesystem-specific subset of the \emph{Linux Test
Project} (LTP) \citep{larson:2002:LTP,ltp-github} benchmark suites
to provoke lock operations for and accesses to VFS data structures.

However, when we actually measure basic-block test coverage in Linux
with kcov \citep{vyukov:2016:kcov} while running all LTP suites that
even remotely seem to be  related to VFS, our results indicate that
only about 35 percent of the kernel's basic blocks associated with
the VFS subsystem are actually executed. Although it is not reasonable
to expect 100 percent coverage -- e.g., depending on the kernel configuration
there may be several filesystems compiled in that are not associated
with an actual mount point on the test system -- there is certainly
room for improvement.

In this article, we propose an approach to increase kernel-code coverage
for LockDoc: We repurpose \emph{syzkaller} \citep{vyukov:2015:syzkaller},
a coverage-guided fuzzer with the goal to validate the robustness
of kernel APIs, to not aim for kernel crashes but only for basic-block
coverage in a particular kernel subsystem. The thereby generated
new benchmark programs can be run in addition to the LTP suites, and
increase VFS basic-block coverage by 26.1 percent from 34.7 percent
(LTP only) to 43.8 percent (LTP + generated benchmarks).

To summarize, the contributions of this article are:
\begin{itemize}
\item A quantitative analysis of the LTP's capability to trigger code execution
in Linux's VFS subsystem (\prettyref{sec:Approach}).
\item An approach to repurpose a coverage-guided kernel fuzzer to generate
benchmark programs that target a particular kernel subsystem and do
not crash the system (\prettyref{sec:Approach}).
\item An evaluation demonstrating that the combination of LTP and generated
benchmark programs covers significantly more VFS basic blocks than
LTP alone (\prettyref{sec:Evaluation}).
\end{itemize}
\prettyref{sec:Related-Work} discusses related work, and \prettyref{sec:Conclusions}
concludes the paper.

\section{Related Work\label{sec:Related-Work}}

Linux-kernel test coverage has been a research topic since Linux's
very early days. Iyer \citep{iyer:2002:ltp-coverage} analyzes the
LTP's coverage of Linux 2.4 kernel code with GCOV, and reports that
parts of the \texttt{fs/} kernel source-code subtree have line-coverage
values between 0.0 (especially for most of the actual filesystem implementations)
and 29.5 percent (for the generic, filesystem-agnostic part). Larson
\citep{larson:2003:ltp-coverage} distills detailed HTML reports from
GCOV results obtained from Linux 2.5, reports that the LTP covers
about 90 percent of all kernel basic blocks that are executed by a
much larger benchmark corpus, and concludes that coverage results
should drive further LTP development. Yoshioka \citep{yoshioka2007regression}
describes a Linux regression test framework named \emph{crackerjack}
and the accompanying branch-coverage test tool \emph{btrax}, and demonstrates
coverage advantages over LTP. The \emph{Lachesis} approach \citep{claudi:2011:rt-linux-test}
by Claudi and Dragoni provides a test suite focusing on real-time
extensions for Linux.

While OS-kernel fuzzing approaches date back to at least 1991 with
Le's \emph{tsys} \citep{le:1991:tsys} or to 2006 with \emph{trinity}
by Jones et al.\ \citep{jones:2006:trinity}, modern kernel-fuzzing
approaches like Vyukov's \emph{syzkaller} \citep{vyukov:2015:syzkaller}
are coverage-guided. For example, Nossum and Casasnovas \citep{nossum2016filesystem}
port the prominent user-mode fuzzer AFL to the kernel and uncover
filesystem bugs. \emph{DIFUZE} by Corina et al.\ \citep{corina2017difuze}
fuzzes kernel drivers to detect bugs, and is aided by static analysis
that determines the necessary input structure. Schumilo et al.'s \emph{kernel
AFL} (kAFL) \citep{schumilo:2017:kernel-fuzzing} is a target-OS agnostic
fuzzing approach based on a hypervisor and hardware support in the
form of Intel's \emph{processor trace} feature. Shi et al.\ \citep{shi:2019:linux-fuzzing}
demonstrate the practical application of existing kernel-fuzzing tools
to several Linux versions, and provide an overview of several other
kernel-fuzzing approaches.

To the best of our knowledge, the approach described in this paper
is the first to repurpose coverage-guided kernel fuzzing to generate
benchmarks that complement existing test suites.

\section{Approach\label{sec:Approach}}

The quality of LockDoc's results heavily depends on the number of
observed lock operations and data-structure accesses \citep{lochmann:19:lockdoc}.
Due to static-analysis limitations (e.g. pointer aliasing), it is
generally infeasible to statically identify all code locations that
make such accesses. For similar reasons it is generally infeasible
to statically determine all contexts from which these code locations
are called, and consequently, which locks are possibly held when the
data-structure accesses are made. LockDoc therefore resorts to dynamic
analysis, i.e. the observation of the running kernel under a benchmark
load. As LockDoc focuses on the VFS subsystem to determine locking
rules e.g. for the \emph{inode} data structure, the problem at hand
therefore is to find and run benchmarks that maximize kernel-code
coverage for this particular subsystem.

The literature proposes several metrics for code coverage, e.g., path,
branch or line coverage \citep{zhu:1997,myers:2011}. For this paper,
we chose basic-block coverage over line coverage, because it better
captures the actual fraction of the code that is being executed. One
line of code, for example, can be mapped to several basic blocks.
Hence, having one particular line covered does not necessarily mean
all basic blocks are covered. Inversely, covering all basic blocks
means all lines of code that have been compiled are covered.

For the Linux kernel, there is already a good starting point for executing
a large fraction of kernel code: the \emph{Linux Test Project} (LTP)
\citep{larson:2002:LTP,ltp-github}, of which we already used a VFS-related
subset for earlier work on LockDoc \citep{lochmann:19:lockdoc}. LTP's
aim is to \emph{``validate the reliability, robustness, and stability
of Linux''} \citep{ltp-github}. It consists of several individual
tests that are grouped into test suites. Each of these suites targets
a particular subsystem or a particular kernel functionality such as
the IPC mechanism or the VFS subsystem. The scope of the tests ranges
from stress testing to regression testing. Since those tests are manually
composed, only a limited subset of each system call's parameter space
can be covered, resulting in a limited amount of code coverage.

The VFS-related LTP test suites generate a basic-block coverage of
34.7 percent for the VFS subsystem. We determined the coverage for
each test individually of the following test suites\footnote{We use git tag \emph{20190115} of the LTP repository.}:
\emph{dio}, \emph{fcntl-locktests}, \emph{filecaps}, \emph{fs}, \emph{fs\_ext4},
\emph{fs\_perms\_simple}, \emph{fsx}, \emph{io}, and \emph{syscalls}.
As the \emph{fs\_readonly} suite only runs a subset of \emph{fs} on
a read-only mounted filesystem, we skipped it. \prettyref{tab:bb-coverage}
shows the number of VFS basic blocks covered by each suite.

\begin{table}
\begin{tabular}{rSSS[table-format=2.1]}
\toprule 
\textbf{Test Suite} & \textbf{\# Tests} & \textbf{Covered VFS BBs} & \textbf{(\%)}\tabularnewline
\midrule
\midrule 
dio & 30 & 8312 & 11.0\%\tabularnewline
\midrule 
fcntl-locktests & 1 & 2420 & 3.2\%\tabularnewline
\midrule 
filecaps & 1 & 2518 & 3.3\%\tabularnewline
\midrule 
fs & 65 & 17495 & 23.2\%\tabularnewline
\midrule 
fs\_ext4 & 4 & 13081 & 17.3\%\tabularnewline
\midrule 
fs\_perms\_simple & 18 & 5081 & 6.7\%\tabularnewline
\midrule 
fsx & 1 & 6572 & 8.7\%\tabularnewline
\midrule 
io & 2 & 6817 & 9.0\%\tabularnewline
\midrule 
syscalls & 1181 & 24217 & 32.1\%\tabularnewline
\midrule 
Total & 1303 & 26229 & 34.7\%\tabularnewline
\bottomrule
\end{tabular}\caption{\label{tab:bb-coverage}Covered basic blocks for each LTP test suite
in the VFS subsystem. In total, our Linux 4.10 kernel build consists
of 342,732 basic blocks, and the VFS subsystem of 75,531 basic blocks,
respectively.}
\end{table}

The numbers indicate that there is still room for coverage improvement
-- and, hence, higher LockDoc precision and bug-finding effectiveness.
It turns out that \emph{fuzzing} (see \prettyref{sec:Related-Work})
in its coverage-guided feedback variant already has coverage maximization
as one of its main goals. A coverage-guided kernel fuzzer that recently
came to fame is Vyukov's \emph{syzkaller} \citep{vyukov:2015:syzkaller},
which fuzzes the Linux kernel by randomly generating user programs
that use the system-call interface. For each generated program (see
an example in \prettyref{lst:syzkaller-example}), syzkaller determines
the resulting basic-block coverage. Only programs that cover at least
one new basic block are stored in the database. syzkaller's objective
is to trigger kernel bugs, and to minimize the program that triggered
the bug.

\begin{lstlisting}[language={C},float=tbp,breaklines=true,caption={An excerpt of a program generated by fuzzing the Linux Kernel using Syzkaller \citep{vyukov:2015:syzkaller}.},label={lst:syzkaller-example}]
int main(void)
{
  syscall(__NR_mmap, 0x1ffff000, 0x1000, 0, 0x32, -1, 0);
  syscall(__NR_mmap, 0x20000000, 0x1000000, 7, 0x32, -1, 0);
  syscall(__NR_mmap, 0x21000000, 0x1000, 0, 0x32, -1, 0);

  *(uint32_t*)0x20002480 = 0x20000340;
  memcpy((void*)0x20000340, "\x12", 1);
  *(uint32_t*)0x20002484 = 1;
  *(uint32_t*)0x20002488 = 0;
  syz_read_part_table(0, 1, 0x20002480);
  return 0;
}
\end{lstlisting}

We modified syzkaller to 1)~ignore programs triggering a bug, and
to 2)~only store programs that increase the coverage in the VFS subsystem.
We furthermore disabled a set of system calls that are not related
to the VFS to improve fuzzing speed. The resulting programs are intended
to cover more code, and thus cover more memory accesses, which in
turn can be used by LockDoc. Note that these generated benchmark programs
cannot (directly) be used as regression tests for the kernel, as they
do not make any explicit output that can be used to determine test
success. 

\section{Evaluation\label{sec:Evaluation}}

In this section, we first present our evaluation setup in \prettyref{subsec:Setup},
and then show our results in \prettyref{subsec:Results}.

\subsection{Setup\label{subsec:Setup}}

We conduct our experiments on an x86 64-bit Linux Kernel 4.10. The
kernel is built without module support, and uses a minimal kernel
configuration: Network support is active as well as the essential
drivers for the root filesystem and for running in a paravirtualized
\emph{QEMU}-based virtual machine. To record the executed basic blocks,
we enabled a kernel feature called kcov~\citep{vyukov:2016:kcov},
which was initially introduced by syzkaller \citep{vyukov:2015:syzkaller}.

The syzkaller modifications\footnote{Our modifications are based on git commit \emph{056be1b9c8d0c6942412dea4a4a104978a0a9311}.}
mentioned in \prettyref{sec:Approach} include disabling 191 system
calls that are not related to VFS, e.g. those for process control,
memory operations, or network operations. Based on an example given
by syzkaller, we implemented a library that records the executed kernel
basic blocks during execution of an arbitrary program in 307 lines
of \emph{C++} code. Since the library hooks into the program under
test via the \emph{LD\_PRELOAD} mechanism, we collect covered basic
blocks for a complete process hierarchy. We used this library to gather
the results presented in \prettyref{sec:Approach}.

Whether one basic block belongs to the VFS subsystem or not is determined
using \emph{addr2line}\footnote{\url{https://sourceware.org/binutils/docs/binutils/addr2line.html}}
on the Linux-kernel ELF image: It converts an address to one or more
kernel source-file names. Due to function inlining, it may return
more than one source file for a single address. If a source file matches
the following regular expression, a basic block is considered to belong
to the VFS subsystem: \lstinline!/fs/|/mm/|fs\.h|mm\.h!. We also
include the \lstinline!mm! directory and header files containing
\lstinline!mm.h!, because the file-I/O code is located there such
as \lstinline!mm/readahead.c! or \lstinline!mm/page-writeback.c!.

\subsection{Results\label{subsec:Results}}

\begin{figure}
\includegraphics[width=1\columnwidth]{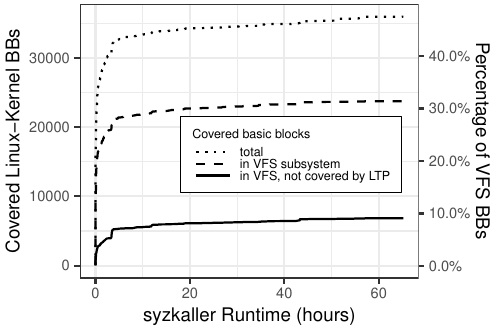}

\caption{\label{fig:syzkaller-coverage-continous}Development of the Linux-kernel
basic-block coverage for the programs generated by syzkaller. The
dotted line shows the number of covered basic blocks out of all 342,732
kernel BBs; the dashed line shows the fraction of the 75,531 basic
blocks of the VFS subsystem. The solid line shows the number of VFS
basic blocks covered by syzkaller-generated programs that are not
already covered by LTP.}

\end{figure}

During its 65-hour run, syzkaller generated 2278 programs that created
a basic-block coverage of 10.0 percent \emph{for the whole kernel}
and 31.4 percent \emph{for the VFS subsystem}. Moreover, it covers
9.1 percent  of VFS basic blocks that are \emph{not already covered
by LTP}. \prettyref{fig:syzkaller-coverage-continous} shows the development
of these three basic-block coverage numbers during syzkaller's run:
The basic-block coverage increases quickly after the start, and slowly
levels off afterwards. The Y axis on the left-hand side of the plot
shows the absolute amount of basic blocks covered, the Y axis on the
right side the percentage of VFS basic blocks. 

\begin{figure}
\includegraphics[width=0.6\columnwidth]{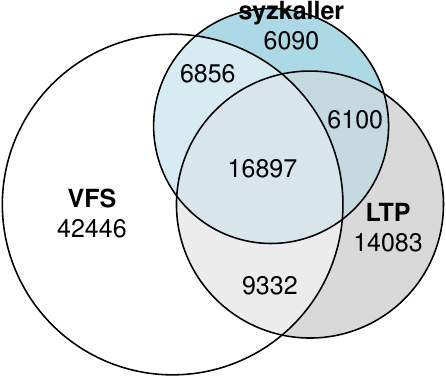}

\caption{\label{fig:bb-coverage-venn}Set relations between VFS-subsystem basic
blocks, and basic blocks covered respectively by syzkaller's generated
programs and by LTP's test suites.}
\end{figure}

The relation between basic blocks belonging to the VFS subsystem,
and those covered by syzkaller's generated programs and by LTP's test
suites, is displayed in the area-equivalent Venn diagram in \prettyref{fig:bb-coverage-venn}.
The absolute numbers shown for each area intersection are the absolute
number of basic blocks shared among all intersecting basic-block sets.

To summarize, our results indicate that combining LTP and syzkaller's
programs to one workload can significantly improve the overall code
coverage for the VFS subsystem by 9.1 percentage points from 34.7
to 43.8 percent.

\section{Conclusions\label{sec:Conclusions}}

In this article, we showed that using LTP as the only benchmark source
for LockDoc yields limited kernel-code coverage, as it only covers
35 percent of basic blocks for the VFS subsystem. We repurposed syzkaller
to generate programs that complement LTP to achieve better coverage,
with the future-work goal to improve LockDoc's precision and bug-finding
effectiveness. We were able to increase the VFS basic-block coverage
by 26.1 percent by combining LTP and syzkaller.

As our next steps, we plan to optimize the resulting benchmark suite
in terms of runtime with near-zero coverage loss. We want to discard
tests or programs that do not add new coverage or incur too much runtime.

\balance

\bibliographystyle{ACM-Reference-Format}
\bibliography{bib/macros-long,bib/all,main,RW/rw}

\end{document}